\documentclass[%
 reprint,
showpacs,preprintnumbers,
 amsmath,amssymb,
 aps,
]{revtex4-1}

\usepackage{graphicx}% Include figure files
\usepackage{dcolumn}% Align table columns on decimal point
\usepackage{bm}% bold math
\newcommand{\be}{\begin{equation}}
\newcommand{\ee}{\end{equation}}
\newcommand{\ba}{\begin{eqnarray}}
\newcommand{\ea}{\end{eqnarray}}
\newcommand{\bd}{\begin{displaymath}}
\newcommand{\ed}{\end{displaymath}}

\begin{document}

\preprint{APS/123-QED}

\title{Shear Viscosities from the Chapman-Enskog and the
Relaxation Time Approaches}% Force line breaks with \\

\author{Anton Wiranata}
\email{awiranata@lbl.gov}
 \affiliation{Institute of Particle Physics, Central China Normal University,
     Wuhan 430079, China\\
     Nuclear Science Division, MS 70R0319 ,Lawrence Berkeley National Laboratory,
     Berkeley, CA 94720, USA}%Lines break automatically or can be forced with \\

\author{Madappa Prakash}%
\email{prakash@harsha.phy.ohiou.edu}
\affiliation{%
Department  of Physics and Astronomy, Ohio University, Athens, OH 45701
}%

\date{\today}% It is always \today, today,

\begin{abstract}
The interpretation of the measured elliptic and higher order collective flows in heavy-ion collisions in terms of viscous hydrodynamics depends sensitively on the ratio of shear viscosity to entropy density.
Here we perform a quantitative comparison between the results of
shear viscosities from the Chapman-Enskog and relaxation time methods
for selected test cases with specified elastic differential cross
sections:
(i) The non-relativistic, relativistic and ultra-relativistic hard sphere gas with
angle and energy independent
differential cross section
(ii) The Maxwell gas, (iii) chiral pions and
(iv) massive pions for which the differential elastic cross section
is taken from experiments.
Our quantitative results reveal that
(i) the extent of agreement (or
disagreement) depends sensitively on the energy dependence of the
differential cross sections employed,
and (ii) stress the need to perform  quantum molecular dynamical (URQMD)
simulations that employ Green-Kubo techniques with similar cross sections  to validate
the codes employed and to test the accuracy of other methods.

\end{abstract}

\maketitle

%\tableofcontents

\newpage
\section{Introduction}

The study of relativistic heavy-ion collisions up to 200 GeV per
particle center of mass energy at the Brookhaven National Laboratory
(BNL), and up to 7~TeV per particle at the Large Hadron
Collider (LHC) at CERN, has required the development of special
theoretical tools to unravel  the complex space-time evolution of the
matter created in these collisions. In view of the large
multiplicities of hadrons (predominantly
pions, kaons, etc.,) observed in these collisions~\cite{PHOBOS:01}, there is much
interest in the description of these collisions from the initial
stages in which quark and gluon degrees of freedom are liberated to
the final stages in which hadrons materialize~\cite{Bjorken:83}.
In a hydrodynamical description of the system's evolution, local thermal
equilibrium is presumed to prevail in the quark-gluon phase, the mixed
phase, and the pure hadronic phase. Thereafter, hadrons cease to
interact (i.e., freeze out) and reach the detectors. Electromagnetic
probes, such as photons and dileptons, produced in matter are expected
to reveal the properties of the dense medium in which they are
produced and from which they escape without any interactions~\cite{PHENIX:09}.  Highly
energetic probes such as jets shed light on the energy loss of quarks
in an interacting dense medium \cite{Gyulassy:90,Wang:92}. In addition, spectral properties
(i.e,. longitudinal and transverse momentum distributions) of the
produced hadrons have revealed interesting collective effects in their
flow patterns~\cite{PHENIX:09}.

A theoretical understanding of the variety of phenomena observed and
expected in these very high energy collisions is clearly a
daunting task.  As a first pass attempt, however, relativistic ideal
hydrodynamics has been fruitfully employed in the description of the
basic facts~\cite{SHKRPV:97,KHHH:01,TLS:01,HKHRV:01}.
Detailed comparisons of the predictions of ideal
hydrodynamics with data have been made, and the merits and demerits of
the theoretical description identified~\cite{BRAHMS:05,PHOBOS:05,STAR:05,PHENIX:05}.
As a result, much attention
has recently been focused on improved developments of viscous relativistic
hydrodynamics.  In addition to the specification of initial conditions
and the knowledge of the equation of state that are the central inputs
to ideal hydrodynamics, the knowledge of transport properties such as
shear and bulk viscosities, diffusion coefficients, etc. is crucial to
viscous hydrodynamics~\cite{RR:07,DT:08,SH:08}.

Our objective in this paper is to quantify the extent to which results from
different approximation schemes for shear viscosities agree (or
disagree) by choosing some classic examples in which the elastic
scattering cross sections are specified. The two different
approximation schemes chosen for this study are the Chapman-Enskog and
the relaxation time methods. These test studies are performed for the following cases:
\begin{enumerate}
\item a hard sphere gas (non-relativistic, relativistic and ultra-relativistic) with  angle and energy independent differential cross
section $\sigma = a^2/4$, where $a$ is the hard sphere radius,

\item  the Maxwell gas ($\sigma (g,\theta)= m\Gamma(\theta)/2g$ with $m$
being the mass of the heat bath particles, $\Gamma(\theta)$ is an
arbitrary function of $\theta$, and $g$ is the relative velocity),

\item chiral pions (for which the $t-$averaged cross section $\sigma =
s/(64\pi^2 f_\pi^4) \left(3 + \cos^2\theta \right)$, where $s$ and $t$ are the usual Mandelstam
variables and $f_\pi$ is the pion-decay constant, and

\item massive pions (for which the differential elastic cross section is taken from
experiments). Where possible, analytical results are obtained in
either the non-relativistic or extremely relativistic cases.
\end{enumerate}

The organization of this paper is as follows. In Sec.~\ref{Formalism},
the formalism and working formulae in the Chapman-Enskog and relaxation time
methods are summarized. Applications to the above mentioned test cases are
considered in  Sec.~\ref{Applications}. A comparison of results from the
two methods is performed on Sec.~\ref{Discussion}.  Our results are summarized in Sec.~\ref{Summary}, which also contains our conclusions. The appendix contains some
details regarding the collision frequency in the non-relativistic limit.

A partial account of this work was given at the International Conference on Critical Point and Onset of Deconfinement held in Wuhan, China, Nov 10, 2011.

\section{Formalism}
\label{Formalism}
In this section, formalisms used to calculate  shear
viscosity using elastic cross sections is described.
In the nonrelativistic regime (such as
encountered in atomic and molecular systems), classic works can be
found in Refs. \cite{Chapman:70, Hirschfelder:54}. Elementary
discussions can be found in Refs. \cite{Reif:1965,McLennan:89}. In the
relativistic regime (as found in cosmology, many astrophysical settings and
relativistic heavy-ion collisions), the book on Relativistic
Kinematics by de Groot \cite{Groot:80} serves as a good reference. For
performing quantitative calculations, the original articles referred
to in this book are more useful. The relevant articles will be
referred to as and when necessary.

In heavy-ion physics, particles of varying masses are produced the
predominant ones being pions (of mass $\sim 140$ MeV), kaons (of mass
$\sim 500$ MeV), etc., the probabilities decreasing with increasing
mass due to energetic considerations. Heavier mass mesons (and baryons
and anti-baryons with masses in excess of the nucleon $\sim 940$ MeV) up to
5 GeV are also produced, albeit in relatively smaller abundances than pions
and kaons. The system is thus a mixture of varying masses evolving in
time from a high temperature (say in the range 200 - 500 MeV) at formation to
100-  150 MeV at freezeout. Thus varying degrees of relativity (gauged in
terms of the individual relativity parameters $z_i=m_i/T$) are
encountered in the mixture. This situation, of varying relativity, is
special to heavy-ion physics. Thus, a general formalism capable of
handling a mixture with varying relativity parameters in time as the system expands is
necessary. In this section, formalisms that address a one-component system in
which particles undergo elastic processes only will be summarized.

It must be stressed that the formalisms used in this work are not new,
but the application of these formalisms to test cases is
new to the extent that a detailed comparison between
two commonly used methods is provided.
For the sake of clarity and
completeness, the formalisms used in this work are summarized below
along with working formulae. This section thus sets the stage for the
ensuing sections in which applications relevant for
 heavy-ion physics will be considered.

\subsection{The Chapman-Enskog Approximation}
\label{chapman}
In this section, the formalism as developed in Ref. \cite{LPG:73} is
followed and described to reveal the essentials.  We begin with the
relativistic transport equation appropriate for a non-degenerate
system:
\begin{equation}
p_{\alpha} \partial^\alpha f = \int (f^{\prime} f_1^{\prime} - ff_1)
\sigma F\, d\Omega^{\prime} d\omega_1 \, ,
\label{boltzmangeneral1}
\end{equation}
using the following notation: $x_\alpha$ and $p_\alpha$ are the space--time and
energy-momentum four vectors. (Metric: $g_{\alpha\beta} = {\rm diag}
(-1,1,1,1)$). The abbreviations $ f \equiv f(x,p), f^{\prime} \equiv
f^{\prime} (x,p^{\prime}), f_1\equiv f(x,p_1)$ and $ f_1^{\prime} \equiv
f^{\prime} (x,p_1^{\prime})$ denote Lorentz invariant distribution functions.
The differential cross--section $\sigma \equiv \sigma(P,\Theta)$ is defined in
the c.m. frame with $ P = [-(p^\alpha + p_1^\alpha) (p_\alpha + p_{1\alpha}
)]^{1/2}$ as the magnitude of the total four--momentum. The invariant flux is
denoted by $F=[(p_\alpha p_1^\alpha)^2-(mc)^4]^{1/2}$ and $d\Omega^{\prime}$
refers to the angles of ${\vec p\prime}$ in the c.m. frame and $d\omega_1 =
d^3p_1/p_1^0$.

For a situation not too far from equilibrium, one may write
\begin{eqnarray}
f=f^0(1+\phi)
\label{chapmandistribution}
\end{eqnarray}
where the deviation function $|\phi| \ll 1$ and $f^0$ is the
Boltzmann distribution function for local
equilibrium~\footnote{The generalization to Bose-Einstein and
Fermi-Dirac Statistics can be found in Ref. \cite{Davasne:96}.}:
\begin{equation}
f^ 0 = \rho z \exp (U_\alpha p^\alpha /kT) / [4\pi(mc)^3K_2 (z)] \, ,
\end{equation}
where, $\rho \equiv \rho (x)$ and $T \equiv T(x)$ are the
particle--number density and temperature in a proper coordinate
system, $U \equiv U(x)$ is the four--velocity of the hydrodynamic
particle flux $(U_\alpha U^\alpha = -c^2)$, and $K_2(z)$ is the modified Bessel
function with $z = mc^2/kT$.  In the first Chapman--Enskog
approximation, the function $\phi (x,p)$ satisfies the equation
\begin{equation}
p_\alpha \partial^\alpha f^0 = - f^0 \mathcal{L}[\phi] \, ,
\label{boltzman2}
\end{equation}
where, $\mathcal{L}[\phi]$ is the linearized collision integral and is given by
\begin{equation}
\mathcal{L}[\phi] = \int \, f_1^0
(\phi +\phi_1 - \phi^{\prime} -\phi_1^{\prime} ) \sigma  F \,
d\Omega^{\prime} d\omega_1 \, .
\end{equation}
The solution to Eq.~(\ref{boltzmangeneral1}) has the general structure
\begin{eqnarray}
\phi &=& A \partial_\alpha U^\alpha -
B \Delta_{\alpha\beta} p^\beta \Delta^{\alpha\beta}
(T^{-1}\partial_\beta T + c^{-2}DU_\beta ) \nonumber \\
&& +
C \langle p_\alpha p_\beta \rangle \langle \partial^\alpha U^\beta \rangle \, ,
\end{eqnarray}
where, the notations $D \equiv U^\alpha\partial_\alpha$,
$\Delta^{\alpha\beta}= g^{\alpha\beta} + c^{-2} U^\alpha U^\beta $, $
\langle t_{\alpha\beta}\rangle= \Delta_{\alpha\beta\gamma\delta}
\,t^{\gamma\delta } $ and $ \Delta _{\alpha\beta\gamma\delta }=
(\Delta_{\alpha\gamma} \Delta _{\beta\delta} + \Delta _{\alpha\delta }
\Delta_{\beta\gamma} )/2 - \Delta_{\alpha\beta }
\Delta_{\gamma\delta}/3 $ have been used.  The scalar functions $A,B$
and $C$, which depend on $p_\alpha U^\alpha (x), \rho (x), U^\alpha
(x)$ and $T (x)$, obey the integral equations
\begin{eqnarray}
\mathcal{L}[A] &=& (-1/ kT) Q \\
\mathcal{L}[B\Delta_{\alpha\beta}\,p^\beta] &=& (1/ kT) (p_\gamma U^\gamma +mh)
\Delta_{\alpha\beta}\, p^{\beta} \\
\mathcal{L}[C\langle p_{\alpha} p_{\beta}\rangle ]
&=& (-1 / kT) \langle p_{\alpha} p_{\beta}\rangle \, ,
\label{linearshear}
\end{eqnarray}
where,
\begin{eqnarray}
Q &=& - (mc)^2/3 + c^{-2}p_\alpha U^\alpha [(1-\gamma )mh+\gamma kT] \nonumber \\
&& +c^{-2}[(4/3) - \gamma ](p_\alpha U^\alpha )^2 \, .
\label{Qforsingle}
\end{eqnarray}
Above, $\gamma = c_p/c_v$ is the ratio of specific heats, and $h =
c^2K_3(z)/K_2(z)$ is the enthalpy at equilibrium. The energy momentum tensor
\begin{equation}
T^{\alpha\beta} \equiv c \int p^{\alpha}p^{\beta}~f~\frac{d^3p}{p^o}
\label{stressenergy}
\end{equation}
can now be calculated with $f = f^o(1+\phi)$. In addition to the
equilibrium energy momentum tensor, the result features terms
involving energy flow and the viscous pressure tensor, which are
defined as
\begin{eqnarray}
 I_q^{\alpha}  \equiv  - U^{\beta}\,T_{\beta\gamma}\,\Delta^{\gamma\alpha}\,,\, \quad
 \Pi^{\alpha\beta} \equiv P^{\alpha\beta} - p\,\Delta^{\alpha\beta}\,\,,
\label{viscouspressure}
\end{eqnarray}
where $P^{\alpha\beta} $ is the pressure tensor defined as $P^{\alpha\beta} \equiv \Delta^{\alpha\gamma}
T_{\gamma\epsilon}\Delta^{\epsilon\beta}$. By employing Eq. (\ref{chapmandistribution}) in
Eqs. (\ref{stressenergy}) and (\ref{viscouspressure}), one can get
\begin{eqnarray}
I_q^{\alpha} & = & -\lambda \Delta^{\alpha\beta}(\partial_{\beta} T
+ c^{_2}T\textnormal{D}U_{\beta})\,, \\
\Pi^{\alpha\beta} &=& -2\eta
\left\langle \partial^{\alpha}U^{\beta}\right\rangle -
\eta_v\Delta^{\alpha\beta}\partial_{\gamma}U^{\gamma} \,.
\end{eqnarray}
The shear viscosity $(\eta_s )$, is given by
\begin{eqnarray}
\eta_{s} &=& -\frac {1}{10} c \int C \langle p_\alpha p_\beta \rangle
 \langle p^\alpha p^\beta \rangle f^{0}\, d\omega \,.
\end{eqnarray}
The above inhomogeneous integral equations for the transport
coefficients can be reduced to sets of algebraic equations by
expanding the unknown scalar function
$C(\tau )$, where $\tau = -(p_\alpha U^\alpha +mc^2)/kT$, in terms of
orthogonal polynomials, e.g. the Laguerre functions $L_n^\alpha (\tau
)$ with appropriate values of $\alpha$ (half integers (0) for massive
(massless) particles).

\subsubsection{Shear Viscosity of a One-Component Gas}
\label{shearonecomponent}
Beginning with Eq. (\ref{linearshear})
\begin{eqnarray}
 \mathcal{L}[C\langle p_{\alpha} p_{\beta}\rangle ]
&=& (-1 / kT) \langle p_{\alpha} p_{\beta}\rangle \,,
\label{shearlinear2}
\end{eqnarray}
the first approximation to the shear viscosity can be obtained
explicitly by (i) multiplying both
sides of the above equation with
\begin{eqnarray}
 \left(p^{(0)}\,f^{(0)}\,L_n^{5/2}(\tau)\, \right)^{-1}\,
\left\langle p^{\alpha}p^{\beta}\right\rangle \nonumber
\end{eqnarray}
and integrating over momentum, (ii) introducing the quantity $\gamma_n$
defined by
\begin{equation}
\gamma_n \equiv \frac{c}{\rho k^2 T^2}\int f^{(0)} L_n^{5/2}(\tau)\left\langle p_{\alpha}p_{\beta}\right\rangle \left\langle p^{\alpha}p^{\beta}\right\rangle \frac{d^3p}{p^0}~,
\label{gamman}
\end{equation}
and applying $\gamma_n$ to Eq. (\ref{shearlinear2}),
 and writing the results in terms of the bracket expression as
\begin{equation}
\left[ C\left\langle p_{\alpha}p_{\beta}\right\rangle, L_n^{5/2}(\tau)\left\langle p^{\alpha}p^{\beta}\right\rangle  \right] = \frac{m^2 k T}{\rho} \gamma_n \qquad (n = 0,1,\cdots)~.
\label{braket2}
\end{equation}
We now write $C$ as an expansion involving the generalized Laguerre
polynomial as
\begin{equation}
C(\tau) = \sum_{m=0}^{\infty} c_m L_m^{5/2}(\tau)
\label{cexpan}
\end{equation}
so that  Eq. (\ref{braket2}) can be written as
\begin{equation}
\sum_{m=0}^{\infty} c_m c_{mn} = \frac{1}{\rho k T}\gamma_n\qquad (n=0,1,\cdots)~,
\label{sumofcmn}
\end{equation}
where
\begin{eqnarray}
c_{mn} &=& \frac{1}{(mkT)^2} \left[ L_m^{5/2}(\tau)\left\langle p_{\alpha}p_{\beta}\right\rangle , L_n^{5/2}(\tau) \left\langle p^{\alpha}p^{\beta}\right\rangle \right]\\
&&  (m,n = 0,1,\cdots)~. \nonumber
\end{eqnarray}
Note that  $c_{mn} = c_{nm}$. The $r$th approximation to the coefficient
$c_m^{(r)}$ is obtained by truncating the sum in Eq. (\ref{sumofcmn}) to $r$
terms; that is,
\begin{equation}
\sum_{m=0}^{r-1} c_m^{(r)} c_{mn} = \frac{1}{\rho k T}\gamma_n \qquad (n = 0,1,\cdots, r-1)~.
\label{sumrel}
\end{equation}
Finally,  the shear viscosity can be written as
\begin{equation}
\eta = \frac{1}{10}(kT)^2 \sum_{m=0}^{\infty} c_m \gamma_m~.
\label{shear2}
\end{equation}
The first, second and third approximations to shear viscosity are
\begin{eqnarray}
 [\eta_s]_1 &=&\frac{1}{10}\,kT\,\frac{\gamma_0^2}{c_{00}}
\label{shearfirst}\\
\left[ \eta_s\right]_2 &=& \frac{1}{10}\,kT\,
\frac{\gamma_0^2\,c_{00} - 2\,\gamma_0\,\gamma_1\,c_{01} + \gamma_1^2\,c_{00}}
{c_{00}c_{11}-c_{01}^2} \label{shear2nd}\\
\left[ \eta_s\right]_3 &=& \frac{\rho(kT)^2}{10}\,\left(c_0\,\gamma_0 + c_1\,\gamma_1 +
c_2\,\gamma_2 \right)\,\,\label{shear3rd},
\end{eqnarray}
where
\begin{eqnarray}
 \gamma_0 &=& -10 \hat{h} \\
 \gamma_1 &=& - \left[ \hat{h}(10z -25) -10z \right] \\
 c_{00} &=& 16\left( w_2^{(2)} - \frac{1}{z}\,w_1^{(2)} + \frac{1}{3z^2}w_0^{(2)} \right) \label{c00massive}\\
c_{01} &=& 8 \left(2z\left(w_2^{(2)} - w_3^{(2)} \right) + \left( -2w_1^{(2)} + 3w_2^{(2)}\right)\right. \nonumber \\
&& \left. + z^{-1}\left(\frac{2}{3}w_0^{(2)} - 9w_1^{(2)} \right) -\frac{11}{3z^2}w_0^{(2)} \right)\nonumber \\
\\
c_{11} &=& 4\left(4z^2 \left(w_2^{(2)}-2w_3^{(2)} + w_4^{(2)}\right) \right. \nonumber \\
&& \left. + 2z \left(
-2w_1^{(2)}+ 6w_2^{(2)} - 9w_3^{(2)} \right) \right. \nonumber \\
&& \left. + \left(\frac{4}{3}w_0^{(2)} -36w_1^{(2)} +41w_2^{(2)} \right) \right. \nonumber \\
&& \left. + z^{-1}
\left(-\frac{44}{3}w_0^{(2)} - 35w_1^{(2)} \right) + \frac{175}{3z^2}w_0^{(2)}  \right)
\label{shearcoeffs}\nonumber \\
\end{eqnarray}
with
\begin{eqnarray}
z = \frac{mc^2}{kT}\quad {\rm and} \quad \hat{h} = \frac{K_3(z)}{K_2(z)}
\end{eqnarray}
and the quantity $w_i^{(s)}$ is so-called the relativistic omega integral which is defined as
\begin{eqnarray}
w_{i}^{(s)} &=& \frac{2\pi z^3c}{K_2(z)^2}\int_{0}^{\infty} d\psi
\sinh^7  \psi \cosh^i\psi K_j(2z\cosh\psi)\nonumber\\
& & \times\,\int_{0}^{\pi} d\Theta \sin \Theta \sigma(\psi,\Theta)(1-\cos^s\Theta)~.
\label{omega1}
\label{relomega}
\end{eqnarray}
In the third
order calculation, one more equation is needed to get the relation
between the coefficients $c_n$ and the coefficients $c_{mn}$ which is
shown in Eq. (\ref{sumrel}). The quantity $\sigma(\psi,\Theta)$ is the transport cross section and
$j = \frac{5}{3}+\frac{1}{2}\left( -1\right)^i$;  the others symbols are :
\begin{eqnarray}
g &=& \frac{1}{2}(p_1-p_2)\quad\textnormal{and} \quad P =
(-p_{\alpha}p^{\alpha})^{1/2}\\
\sinh \psi &=& \frac{g}{mc}\qquad\textnormal{ and } \qquad
\cosh \psi = \frac{P}{2mc}~.
\label{omcoeffs}
\end{eqnarray}

\subsubsection{Massless Particles}

For nearly massless particles such as neutrinos and light quarks for which $m/T
\rightarrow 0$, the formalism described earlier can be simplified as
discussed in Ref. \cite{deGroot:75} and is summarized below. The reason
for addressing the ultra-relativistic case is twofold: (1) For
temperatures such that $z_i=m_i/T \rightarrow 0$, as is the case for
light quarks in the context of heavy-ion collisions, it serves as a
first orientation toward the magitudes of viscosities, and (2) test
cases for validating Green-Kubo calculations can be set up in this
limit.

We start again with the relativistic transport equation for a
one-component system of nondegenerate particles:
\begin{eqnarray}
 p^{\alpha}\,\partial_{\alpha}\,f(x,P) = \int\, \left(f\,f_1 - f{'}\,f_1{'}
\right)
\sigma\,F\,d\Omega{'}\,dw_1\,\,,
\label{boltzmangeneral}
\end{eqnarray}
where $f = f(x,p)$, $\sigma = \sigma(\Theta, P)$ is the scattering
cross section for $ p + p_1 \rightarrow p{'} + p_1{'}$ in the center
of momentum frame. Other symbols are
\begin{eqnarray}
 F &\equiv& \left[ (p^{\alpha}p_{1\alpha})^2 - p^{\alpha}p_{\alpha} p^{1\beta}p_{1\beta}\right]^{1/2} \nonumber \\
&=& -p^{\alpha}p_{1\alpha} = \frac{1}{2}P \\
d\Omega{'} &\equiv& \sin\theta{'}\,d\theta{'}\,d\phi{'} \\
dw_1 &\equiv& \frac{d^3p_1}{p_1^0}\,\,,
\end{eqnarray}
where $\theta{'}$ and $\phi{'}$ are the polar angles of the three momentum $\vec{p}\prime$ in the
center of mass frame.

For massless particles,
the equilibrium distribution function can be written as
\begin{eqnarray}
 f_{{\rm eq}} = \frac{nc^3}{8\,\pi\,(k_BT)^3}\,\exp\left[ p^{\alpha}U_{\alpha}/(k_BT) \right]\,\,,
\end{eqnarray}
where $n$ is the number density of particles, $c$ is the speed of light, $k$ is the
Boltzman constant, $T$ is the temperature and $U$ is the flow velocity.
In the first Enskog approximation, the perturbed distribution function of the system can be
written as
\begin{eqnarray}
 f(x,p) = f^{(0)}(x,p) \left[1+ \phi(x,p) \right]\,\,,
\label{distmassless}
\end{eqnarray}
where $f^{(0)}(x,p)$ is the local equilibrium distribution function and $\phi(x,p)$ is
the deviation function. Using Eq. (\ref{distmassless}), one can linearize Eq. (\ref{boltzmangeneral})
to get
\begin{eqnarray}
 \left( p^{\alpha}U_{\alpha} + 4k_BT \right)\,p^{\beta}\,\Delta_{\beta\gamma} \left( T^{-1}
\partial^{\gamma} T + c^{-2}D\,U^{\gamma} \right)  \nonumber \\
+ \left\langle p^{\alpha}p^{\beta} \right\rangle + \left\langle \partial_{\alpha}U{\beta} \right\rangle = -k_BT\mathcal{L}\left[\phi \right]\,\,,
\label{boltzmanlinear}
\end{eqnarray}
where $\mathcal{L}$ is the linearized operator defined by
\begin{eqnarray}
 \mathcal{L} \equiv \frac{1}{2}\,\int\,f_1^{(0)}\,\delta(F)\,\sigma\,P^2\,d\Omega{'}\,dw_1
\end{eqnarray}
with
\begin{eqnarray}
 \delta(F) \equiv F(p) + F(p_1) - F(p{'}) -F(p_1{'})\,\,.
\end{eqnarray}
In Eq. (\ref{boltzmanlinear}), the angular bracket $\left\langle \cdots \right\rangle$ is
for the operation
\begin{eqnarray}
 \left\langle A_{\alpha\beta} \right\rangle \equiv \frac{1}{2}\,\Delta_{\alpha}^{\beta}\,
\left( A_{\gamma\delta}+A_{\delta\gamma} \right)\Delta_{\beta}^{\delta} -
\frac{1}{3}\,\Delta_{\alpha\beta}\,\Delta_{\gamma\delta}\,A^{\gamma\delta}\,\,.
\end{eqnarray}
The general form of the deviation function (for elastic collisions) is
\begin{eqnarray}
 \phi(x,p) &=& -B_{\alpha} \Delta^{\alpha\beta}\,\left( T^{-1}\partial_{\beta}T +
c^{-2}DU_{\beta}\right) \nonumber \\
&&+ C_{\alpha\beta} \left\langle \partial^{\alpha}U^{\beta} \right\rangle \,.
\end{eqnarray}
In the case of shear viscosity, one needs to solve for the coefficients $C_{\alpha\beta}$
which satisfy
\begin{eqnarray}
 \mathcal{L} \left[C_{\alpha\beta} \right] = - (kT)^{-1}\,\left\langle p_{\alpha}p_{\beta}
\right\rangle\,\,.
\label{linear2}
\end{eqnarray}
In order to get an expression for the shear viscosity, one can use the distribution function
in Eq. (\ref{distmassless}) in the viscous pressure tensor which is defined as
\begin{eqnarray}
 \pi^{\alpha\beta} \equiv P^{\alpha\beta} - p\Delta^{\alpha\beta}\,,\quad {\rm where} \quad
P^{\alpha\beta} \equiv \Delta^{\alpha\beta}T_{\gamma\delta}\Delta^{\delta\beta}\,\,.
\end{eqnarray}
As a result,
\begin{eqnarray}
 \pi^{\alpha\beta} = -2\,\eta_s\left\langle \partial^{\alpha}U^{\beta} \right\rangle \,\,,
\end{eqnarray}
where
\begin{eqnarray}
 \eta_s \equiv \frac{1}{10}\,c\,\int \,C_{\alpha\beta}
\left\langle p_{\alpha}p_{\beta} \right\rangle \,f^{(0)} \,dw\,\,,
\end{eqnarray}
with $C_{\alpha\beta} = C\,\left\langle p^{\alpha}p^{\beta} \right\rangle$. The coefficient
$C$ can be written in terms of associated Laguerre polynomials:
\begin{eqnarray}
 C(\tau) = \sum_{n=0}^{\infty}\,c_n\,L_n^5(\tau)\,\,,
\label{coeffcmassless}
\end{eqnarray}
where
\begin{eqnarray}
 L_n^{\alpha}(\tau) \equiv \sum_{i=1}^n\,
\binom{n+\alpha}{i+\alpha}\,
\,\frac{(-\tau)^i}{i!}\,, \quad {\rm where} \quad
\tau \equiv -p^{\alpha}U_{\alpha}/kT\,. \nonumber \\
\end{eqnarray}
These functions satisfy the relations
\begin{eqnarray}
\int_0^{\infty} L_m^{\alpha}(\tau)L_n^{\alpha}(\tau) \,\tau^{\alpha}
\exp(-\tau)\,d\tau = \left[\frac{\Gamma(n+\alpha+1)}{n!} \right]\,\delta_{0m}\,. \nonumber \\
\end{eqnarray}
Inserting Eq. (\ref{coeffcmassless}) in Eq. (\ref{linear2}), one gets
\begin{eqnarray}
 \sum_0^{\infty}\,c_n\,\mathcal{L}\left[ L_n^5(\tau) \left\langle
p_{\alpha}p_{\beta} \right\rangle \right] = -(kT)^{-1} L_)^5(\tau)
\left\langle p_{\alpha}p_{\beta} \right\rangle\,\,.
\label{sumofcnlinear}
\end{eqnarray}
Introducing the notation
\begin{eqnarray}
 c_{mn} \equiv \left[ L_m^5(\tau) \left\langle p^{\alpha}p^{\beta} \right\rangle,
L_n^5(\tau)\left\langle p_{\alpha}p_{\beta} \right\rangle \right] \,,
\end{eqnarray}
where the bracket operation means
\begin{eqnarray}
 \left[F,G \right] &\equiv&
\frac{1}{8}\,cn^{-2}\,\int\,\delta(F)\delta(G)f^{(0)}f_1^{(0)}\sigma\,P^2\,
dw\,dw_1\,d\Omega{'}\,. \nonumber \\
\end{eqnarray}
Equation (\ref{sumofcnlinear}) can be written as
\begin{eqnarray}
 \sum_{n=0}^{r-1}\,c_{mn}\,c_n = -40\,n^{-1}\,c^{-2}\,(kT)^2\,\delta_{0m}\,.
\label{cmns}
\end{eqnarray}
Hence, one can write the shear viscosity for massless particles as
\begin{eqnarray}
 \eta_s = -40 \,nc^{-2}\,(kT)^3\,c_0
\label{shearmassless}
\end{eqnarray}
\begin{table*}
\caption{\label{masslessviscosity}The values of
$\gamma, \epsilon, t, u, v$ and $w$ as a
function of $i, p, q, r$ and $s$ for the case of shear viscosity. }
\begin{tabular}{|c|c|c|c|c|c|c|c|c|}
\hline\hline
i&$\gamma$ & $(-1)^p$ & $(-1)^q$ & $\epsilon$ & $t-p$ & $u-q$ & $2v-p-q-2$ & $w+p+q-r-s$  \\
\hline\hline
1&1&+&+&2&0&0&0&0\\
2&2&-&-&1&0&0&2&0\\
3&2&+&+&1&1&1&2&0\\
4&-2&+&-&1&1&0&1&1\\
5&-2&-&+&1&0&1&1&1\\
6&2/3&+&+&0&0&0&0&4\\
7&-4/3&-&+&0&1&0&1&3\\
8&-4/3&+&-&0&0&1&1&3\\
9&2/3&+&+&0&2&0&2&2\\
10&8/3&-&-&0&1&1&2&2\\
11&2/3&+&+&0&0&2&2&2\\
12&-2&+&+&0&0&0&0&2\\
13&2&-&+&0&1&0&1&1\\
14&2&+&-&0&0&1&1&1\\
15&-4/3&+&-&0&2&1&3&1\\
16&-4/3&-&+&0&1&2&3&1\\
17&2/3&+&+&0&2&2&4&0\\
18&-2&-&-&0&1&2&2&0\\
19&1&+&+&0&0&0&0&0\\
\hline\hline
\end{tabular}
\end{table*}

In Eq. (\ref{cmns}), the coefficients $c_{mn}$ are calculated from
\begin{equation}
 c_{mn} = \sum_{i=1}^{19}\,c_{mn,i}
\end{equation}
with
\begin{eqnarray}
 c_{mn,i} &=& \frac{\gamma}{(2\beta c)^2}\,\sum_{r=0}^m\,\sum_{s=0}^n\,
\binom{m+5}{r+5} \,\binom{n+5}{s+5}\,(-2)^{-r-s} \nonumber \\
&&\times \sum_{p=0}^r{'}\sum_{q=0}^s{'}(-1)^{p+q}\frac{t!}{p!(r-p)!}\frac{u!}{q!(s-q)!}\frac{[(t+u)/2]!}{(t+u+1)!}\nonumber \\
&&\times \sum_{k=0}^{[M/2]}\,\frac{2^{M-2k}}{k!(k+|t-u|/2)!(M-2k)!}\nonumber \\
&&\times \sum_{k=0}^{[w/2]}\,\binom{[w/2]}{0} \,
(2v+2l-1)!!\,\,\tilde{w}_{r+s-l-v+9, l+v+\delta}^{M-2k+\epsilon}\,\,, \nonumber \\
\label{cmnis}
\end{eqnarray}
where $\tilde{w}_{ij}^k$ is the omega integral for massless particles :
\begin{eqnarray}
\tilde{w}_{ij}^k &\equiv& \frac{\pi}{2^4\beta}\,\int_{-1}^1\,d\cos\Theta\,\int_0^{\infty}\,
dP\,\sigma(\Theta,P)\,(1-\cos^k\Theta) \,\nonumber \\
&& \times\, (\beta c P)^i\,K_j(\beta c P) \,.
\label{omegamassless}
\end{eqnarray}

The quantity $\delta$ is given by $\delta \equiv (r+s)$ mod$ (2)$ and
the quantity $M$ is given by $M = \min (t,u)$. The rest of the
variables needed are listed in Table.  \ref{masslessviscosity}, which is reproduced from
Ref. \cite{deGroot:75}  in which details of the derivation that leads to the form shown in Eq. (\ref{cmnis}) for $c_{mn,i}$ are given.

In the first order approximation, the required coefficients are
\begin{eqnarray}
 c_{00} &=& \left( \frac{1}{3}\tilde{w}_{63}^2 + \frac{1}{2}\tilde{w}_{72}^2 + \frac{1}{4}\tilde{w}_{81}^2 \right)/
(\beta c)^2 \label{c00massless}\\
c_{01} &=& \left(2\tilde{w}_{63}^2 + 3\tilde{w}_{72}^2 -\frac{1}{2}\tilde{w}_{74}^2 + \frac{3}{2}\tilde{w}_{81}^2 -
\frac{1}{2}\tilde{w}_{83}^2 \right. \nonumber \\
&& \left. - \frac{1}{8}\tilde{w}_{92}^2 \right) /(\beta c)^2 \\
c_{11} &=& \left( 12\tilde{w}_{63}^2 + 18\tilde{w}_{72}^2 - \frac{3}{4}\tilde{w}_{74}^2 + 9\tilde{w}_{81}^2 -
3\tilde{w}_{83}^2  \right. \nonumber \\
&& \left. - \frac{15}{16}\tilde{w}_{92}^2 + \frac{1}{16}\tilde{w}_{10,1}^2 \right)/(\beta c)^2\,. \nonumber \\
\end{eqnarray}

The scheme outlined above has been utilized to calculate the shear
viscosity of neutrinos in Ref. \cite{deGroot:75} and of chiral
pions by Prakash et al. in \cite{PPVW:93}. In the next section, an
application of this scheme to calculate $\eta_s$ with a constant
cross section  will be presented. This application will be utilized to
validate the Green-Kubo calculations.

\subsubsection*{Deviations from the Ultra-Relativistic Limit}

The ultra-relativistic limit corresponds to the relativity parameter
$z=mc^2/kT \rightarrow 0$, in situations when either the mass tends to
vanish or when the temperature is very large compared to the mass.  In
the context of relativistic heavy ion collisions, low-mass quarks such
as the $u$ and $d$ quarks with current quark masses $\leq 10$ MeV in
conditions of temperatures above the phase transition temperature of
$kT \sim 200$ MeV, fall into the category of $z \ll 1$. In the
hadronic phase, pions of masses $\sim 140$ MeV in the temperature
range of $100-200$ MeV, however fall in the borderline regime of the
intermediate relativistic regime. It is therefore of some interest to
gauge how deviations from the ultra-relativistic regime affect the
transport coefficients. In this section, we summarize the work of Ref.
\cite{Kox:76} in which effects of slight deviations from the
ultra-relativistic case were established in the case of hard
spheres. Thereafter, the formalism for arbitrary interactions is
developed. The case of the hard spheres will serve as a testbed for
calculations of transport coefficients from the Green-Kubo formulas
 in which the mass of the particle is
set to a small value for computational ease. Our development for
arbitrary interactions will further aid the validation of such
calculations in the relativistic regime.

\subsubsection{The Hard Sphere Gas}

Here the calculation of Ref. \cite{Kox:76} for the hard sphere gas with
a constant differential cross section $\sigma_0 = a^2/4$, where $a$ is
the radius of the particle, is summarized.

The first step is to
rewrite  the relativistic omega integral in Eq. (\ref{omega1}),
with $x = \cosh\,\psi$, as
\begin{eqnarray}
 w_i^{(s)} &=& \frac{\pi z^3cf^{(s)}a^2}{2K^2_2(z)}
 \int_1^\infty\,dx\,\left(x^2-1 \right)^3\,x^i\,K_j(2zx) \nonumber \\
\end{eqnarray}
where
\begin{eqnarray}
j = 5/2 + 1/2(-1)^i \quad
{\rm and}\quad
f^{(s)} \equiv \frac{\left[2s+1+(-1)^{s+1} \right]}{(s+1)\,}. \nonumber \\
\end{eqnarray}
By changing the integration variable and by employing binomial coefficients
to express the third power, one can rewrite the above equation as
\begin{eqnarray}
 w_i^{(s)} &=& \left[ \pi\,z^{-i-4}\,c\,f^{(s)}\,a^2 / 2K^2_2(z) \right]\,
\sum_{k=0}^3 \binom{3}{k}\,(-1)^k\,z^{2k}\nonumber \\
 && \times\,\int_z^\infty\,dx\,x^{i-2k+6} \,K_j(2x)\,.
\label{omegaz}
\end{eqnarray}
In the limit of $ z \ll 1 $, the two modified Bessel functions have the
behaviors
\begin{eqnarray}
 K_2(z) &=& \frac{2}{z^{-2}}\,\left[ 1 - \frac{1}{4} z^2 - \frac{1}{16}z^4 \ln\,z
+ \frac{1.731863}{32}\,z^4+ \cdots \right] \label{bessel2}\nonumber \\
\\
 K_3(z) &=&\frac{8}{z^{-3}} \,\left[ 1 - \frac{1}{8} z^2 + \frac{1}{64}\,z^4
+  \cdots \right] \label{bessel3} \,.
\end{eqnarray}
Also, in the limit of $z \rightarrow 0 $, the integral
involving the modified Bessel function can be written as
\begin{eqnarray}
 \int_0^\infty\,dx\,x^\mu\,K_\nu(2x) = \frac{1}{4}\,\Gamma\left[(\mu+\nu+1)/2
\right]\,\Gamma\left[(\mu-\nu+1)/2 \right]\,.\nonumber \\
\end{eqnarray}
The above relation is true for $\mu \pm \nu + 1 > 0$. Using these relations,
the omega integral in Eq. (\ref{omegaz}) reads
as
\begin{eqnarray}
 w_i^{(s)} &=& \frac{1}{32}\,\pi\,z^{-1}\,c\,f^{(s)}\,a^2\,
\Gamma\left[(i +j+7)/2 \right]\,\Gamma\left[(i-j+7)/2 \right]\, \nonumber \\
&& \times \, \left(1 + \frac{(i^2-j^2+10i+1)}{(i^2-j^2+10i+25)}\,
\frac{z^2}{2} + \cdots\right)\,.
\end{eqnarray}
Thermodynamic quantities such as the enthalpy, $h$, and the ratio of
specific heats, $\gamma$, can be evaluated in the $z\rightarrow 0$ limit
by applying the properties of modified Bessel function in
Eqs. (\ref{bessel2}) and (\ref{bessel3}) so as to read as
\begin{eqnarray}
 h &=& 4\,c^2\,z^{-1}\,\left[1 +\frac{1}{8}z^2+\frac{1}{16}z^4\,\ln\,z \right.\nonumber \\
&& \left. + \frac{1}{32}\left(\frac{3}{2} - 1.731863\right)z^4 + \cdots \right] \\
\gamma &=& \frac{4}{3}\,\left[1 +\frac{1}{24}z^2+\frac{1}{64}z^4\,\ln\,z \right. \nonumber \\
&&\left.  + \frac{1}{32}\left(\frac{43}{18} - 1.731862\right)z^4+\cdots \right]\,.
\end{eqnarray}
Then the shear viscosities in this regime read as
\begin{eqnarray}
\left[ \eta_s\right]_p &=& \frac{mc}{\pi\,a^2}\,z^{-1}\,F_p\,
\left(1 + F^{'}_p\,z^2\,+ \cdots \right)\,,
\label{etasz0}
\end{eqnarray}
where the subscript $p$ refers to the $p$th approximation
and the values of the various coefficients in the above equations
are listed in Table. \ref{valueofconstant} for $p=1,2$ and 3.
\begin{table}[h]%The best place to locate the table environment is directly after its first reference in text
\caption{\label{valueofconstant}Values of the coefficients
appearing in Eq. (\ref{etasz0}) for the hard sphere gas.}
\begin{ruledtabular}
\begin{tabular}{lcdr}
\textrm{p}&
\textrm{1}&
\multicolumn{1}{c}{\textrm{2}}&
\textrm{3}\\
\colrule
& & & \\
${\rm F}_p$ & 1.2 & 1.2588 & 1.2642\\
${\rm F}'_p$ & 0.05 & 0.0424 & 0.0403\\
& & & \\
\end{tabular}
\end{ruledtabular}
\end{table}

Note that for massless particles,
\begin{eqnarray}
[\eta_s]_p = \frac {k_BT}{\pi a^2} \frac 1c ~F_p\,.
\end{eqnarray}
In the next section, calculations that attest to
the rapid convergence of the coefficient $F_p$
are carried to much higher order in $p$. These results will be of much
utility in validating ultra-relativistic molecular dynamical
simulations of shear viscosity.

%\clearpage

\subsection*{Reduction to the Non-Relativistic Case}

In the non-relativistic limit, i.e.,  $z =m/k_BT\gg 1$,
the results above can be further simplified.
As for $z>>1$,
\begin{eqnarray}
  K_n(z) &\simeq& \sqrt{\frac{\pi}{2z}} e^{-z}\,\left(1+\frac{(4n^2-1)}{2!}\,\frac{1}{z}+ \cdots \right)\,\,,
  \label{abesk}
\end{eqnarray}
the reduced enthalpy
\be
 \hat{h} \rightarrow 1\qquad  {\rm and} \qquad \gamma_0 \rightarrow -10 \,.
\label{ahhat}
\ee
In addition, the relativistic omega integral, $w_i^{(s)}$ in Eq.~(\ref{relomega}) can be transformed into its non-relativistic counterpart
$\Omega_i^{(s)}$ as follows.
Introducing the dimensionless quantity
\be
 \phi = \frac{g}{\sqrt{mc^2\,k_BT}} = \frac{mc\,\sinh\,\psi}{\sqrt{mc^2\,k_BT}} \,\,,
 \ee
 whereby
\be
\cosh\,\psi = \sqrt{1+z^{-1}\phi^2}\,\,,
\ee
the integral over $\psi$ in Eq.~(\ref{relomega}) can be written as
\begin{eqnarray}
 I_{\psi} &=& \frac{2\,\pi\,k_BT}{m\,K_2^2(z)}\,\int_0^{\infty}\,d\phi\,\,\phi^7\,(1+z^{-1}\phi^2)^{i-1}~\nonumber \\
&& \times  \,K_j(2z\sqrt{1+z^{-1}\phi^2})\,.
\label{Ipsi}
\end{eqnarray}
The use of a binomial expansion and the expansion of $K_n(z)$ for $z \gg 1$,
\begin{eqnarray}
  K_n(z) &\simeq& \sqrt{\frac{\pi}{2z}} e^{-z}\,\left(1+\frac{(4n^2-1)}{2!}\,\frac{1}{z}+ \cdots \right)\,\,,
  \label{abesk2}
\end{eqnarray}
reduces the above integral to
\begin{eqnarray}
 I_{\psi} = 2\,\sqrt{\frac{\pi\,k_BT}{m}} \, \int_0^{\infty}\,d\phi\,\,\phi^7\,e^{-\phi^2} \,.
\end{eqnarray}
Inserting this result in Eq.~(\ref{relomega}), the omega integral in the non-relativistic limit is
\begin{eqnarray}
 \Omega_2^{(s)} &=& 2\sqrt{\frac{\pi\,k_BT}{m}}\,\, \int d\phi\,\,\phi^7\,e^{-\phi^2} \nonumber \\
&\times& \int_0^{\pi}\, d\theta\, \sin \theta\,\sigma(\phi,\theta)\, (1-\cos^s\theta) \,.
 \label{nrelomega}
\end{eqnarray}

Note that the magnitudes of the omega integrals in
Eqs.~(\ref{relomega}) and (\ref{nrelomega}) are determined by a
combination of different physical factors: the thermal weight, collisions with
large relative momenta, and the relative momentum dependence of the
transport cross section. These omega integrals also feature in the
calculation of the shear viscosity in higher order formulations;
expressions for viscosity in higher order approximations may be found in
Ref.~\cite{LPG:73}.

A further simplification occurs in the non-relativistic limit with
\be
c_{00}  \approx 16\, w_2^{(2)} \rightarrow  16\,\Omega_2^{(2)} \,,
\ee
as the second and third terms in Eq.~(\ref{c00massive}) are suppressed by $z$ being large.  Thus, the shear viscosity takes the simple form
\be
\eta_s = \frac{5}{8}~\frac {k_BT}{\Omega_2^{(2)}}\,.
\label{etasCE}
\ee
\subsection{The Relaxation Time Approximation}

In the relaxation time approximation, the main assumption is that the
effect of collisions is always to bring the perturbed distribution
function close to the equilibrium distribution function, that is $f({\bf x},{\bf p})
\rightarrow f^{(0)}({\bf x},{\bf p})$. In other words,
the effect of collisions is to restore the
local equilibrium distribution function exponentialy with a relaxation
time $\tau_0$ which is of order the time required between particle
collisions \cite{Reif:1965}:
\begin{eqnarray}
 D_c f({\bf x},{\bf p}) = - \frac{f({\bf x},{\bf p}) - f^{(0)}({\bf x},{\bf p})}{\tau_0}
\label{collisionterm}
\end{eqnarray}

In the relativistic case,
we follow closely the formalism described in the review article
by Kapusta~\cite{Kapusta:09} and develop working formulae for the
calculation of shear viscosity using Maxwell Boltzmann
statistics~\cite{Gavin:85,Kapusta:09,CK:10}. (Bose-Einstein and
Fermi-Dirac cases will be considered later.) We restrict our attention
to the case involving two-body elastic reactions $a+b \rightarrow c+d$
in a heat bath containing a single species of particles. In what follows,
we use the notation employed in Ref. \cite{CK:10}. Differences from earlier
notation in this chapter are small, and should not cause any confusion.

In the relaxation time approximation, the shear viscosity is given by
\cite{CK:10}
\begin{eqnarray}
 \eta_s &=& \frac{1}{15T}\,\int_0^{\infty}\,\frac{d^3p_a}{(2\pi)^3}\,
\frac{|p_a|^4}{E_a^2}\,\frac{1}{w_a(E_a)}\,f^{eq}_a\,\,
\label{etasrelax}
\end{eqnarray}
where $w_a(E_a)$ is the collision frequency and $f_a^{eq}$ is the
equilibrium distribution function of particles $a$ with momenta $p_a$
and energy $E_a$:
\begin{equation}
f_a({\bf x},{\bf p}_a,t) = \frac{1}{{\rm e}^{(E_a-\mu_a)/T} - (-1)^{2s_a}} \,,
\end{equation}
where $\mu_a$ is the chemical potential of the particle, and $f_a$ is
normalized such that integration over momenta yields the density
$n({\bf x},t)$.

The collision frequency is given by
\begin{equation}
w_a(E_a) = \sum_{bcd} \frac{1}{2} \int \frac{d^3p_b}{(2\pi)^3} \,
\frac{d^3p_c}{(2\pi)^3} \,
\frac{d^3p_d}{(2\pi)^3} \, W(a,b|c,d)\, f^{eq}_b\,\,,
\label{collisionfrequency}
\end{equation}
where the quantity $W(a,b|c,d)$ is defined as
\begin{equation}
 W(a,b|c,d) = \frac{(2\pi)^4 \delta^4\,(p_a+p_b-p_c-p_d)}
{2E_a2E_b2E_c2E_d} \,|\mathcal{M}|^2\,\,.
\end{equation}
Above, $|\mathcal{M}|^2$ is the squared transition amplitude for the 2-body
reaction $a+b \rightarrow c+d$ and
 $f_b^{eq}$ is the distribution function of particles $b$.
Utilizing the above expression, one can write
\begin{equation}
 w_a(E_a) = \sum_{bcd} \frac{1}{2} \int \frac{d^3p_b}{(2\pi)^3} \,\, f^{eq}_b\,\,I_E\,\,,
 \label{relax1}
\end{equation}
where
\begin{equation}
 I_E \equiv \int \frac{d^3p_c}{(2\pi)^3}\,\frac{d^3p_d}{(2\pi)^3}\,\frac{(2\pi)^4\delta^4(p_a+p_b-p_c-p_d)}{2E_c\,2E_d}\,|\mathcal{M} |^2 \,.
\end{equation}
The exit channel integrals in the above equation
can be manipulated in the center of mass
(c.m.) frame to feature the differential
cross section.
In the c.m. frame, $ \sqrt{s} = E_{cm} = E_c^{cm} + E_d^{cm} = 2E_{c}^{cm} = 2E_{d}^{cm} = 2\,(m^2 + q_{cm}^2)^{1/2}$.
Performing the integration over $\vec{p}{_d}$ in the c.m. frame,
\begin{eqnarray}
 I_E &=& \int \frac{d^3p_c}{(2\pi)^32E_{c}^{cm}} \,\frac{(2\pi)\, \delta\,(\sqrt{s}-E_{cm})}{2E_{c}^{cm}}\, \,|\mathcal{M}|^2 \,.
\end{eqnarray}
The integration over
$p_c$ can be effected through its connection to $q_{cm}$ and $E_{cm}$:
\begin{eqnarray}
q_{cm}^2 &=& \frac{E^2_{cm}}{4} - m^2
\end{eqnarray}
so that $I_E$ can be rewritten as
\begin{eqnarray}
 I_E &=& 2\,\sqrt{s(s - 4m^2)}\,\int\,d\Omega\,\sigma(\Omega)\,\,,
\end{eqnarray}
where
\begin{equation}
 \sigma(\Omega) = \frac{1}{64\,\pi^2\,s}\, |\mathcal{M}|^2
\end{equation}
is the differential cross section in the c.m. frame.
The collision frequency in Eq.~(\ref{collisionfrequency}) thus takes the form
\begin{eqnarray}
  w_a(E_a)  &=&  \int \frac{d^3p_b}{(2\pi)^3}\, \,\frac{\sqrt{s(s-4m^2)}}{2E_a\,2E_b} \,f^{eq}_b\,\sigma_T\,\,,
  \label{freqrelax2}
\end{eqnarray}
where $\sigma_T$ is the total cross section.

As we can see from the above equation, interactions appear in the
collision frequency through the total cross section.  Here we see the
difference with the Chapman-Enskog approximation which features a
transport cross section that favors right-angled collisions in the
c.m. frame.  We shall see that this difference is at the root of
differences in results between the two approaches  in the following sections.
\subsection*{Reduction to the Non-Relativistic Case}
We turn now to reduce Eq.~(\ref{freqrelax2}) for non-relativistic particles. Recalling that $s = 4(m^2+q_{cm}^2)$,
\begin{eqnarray}
 \sqrt{s(s-4m^2)} &=& \sqrt{4(m^2 + q_{cm}^2)\,4q_{cm}^2} \approx 4\,m\,|\vec{q}_{cm}|\,\,\,\,
\end{eqnarray}
in the non-relativistic limit. Also, $E_a \approx m$ and $E_b \approx m$ for equal mass particles. Thus,
\begin{eqnarray}
  w_a(E_a) &=&   \int \frac{d^3p_b}{(2\pi)^3}\,\, \sigma_T\,  f^{eq}_b \, |\vec{v}_a - \vec{v}_b| \,,
  \label{relax3}
\end{eqnarray}
which can be written in the form found in text books (see, e.g., Ref.~\cite{McLennan:89}),
\begin{eqnarray}
 w_a(v_a) &=& \int_0^{\infty} \, d^3v_b\,\sigma_T\,f_b^{eq}\,|\vec{v}_a - \vec{v}_b|
 \label{relax4}
\end{eqnarray}
after a suitable change of variables and normalization of $f_b^{eq}$.

Finally, employing the non-relativistic expressions
$ p_a = m\,v_a$ and $ E_a = m + p_a^2/(2m)$, the shear viscosity
in Eq.~(\ref{etasrelax}) takes the form
\begin{eqnarray}
 \eta_s &=& \frac{1}{30\pi^2}\,\frac{m^5}{T}\,\int\, dv_a\,v_a^6\,\frac{f_a}{w_a(v_a)}
 \label{nretasrelax}
\end{eqnarray}
\section{Applications}
\label{Applications}
\subsection{The Hard Sphere Gas (Non-Relativistic)}

This system is characterized by a constant differential cross section,
$\sigma_0 = a^2/4$, where $a$ is the hard sphere radius.

\subsubsection*{The Chapman - Enskog Approximation}

Utilizing the hard spheres cross section,
the non-relativistic omega integral in
Eq.~(\ref{nrelomega}) becomes
\begin{eqnarray}
 \Omega_2^{(2)} &=& 8\,\,\sigma_0\,\, \sqrt{\frac{\pi\,k_B T}{m}}\,
\end{eqnarray}
Use of this result in Eq.~(\ref{etasCE}) yields
the shear viscosity
\begin{eqnarray}
 \eta_s = \frac{5}{64}\,\sqrt{\frac{m\,k_BT}{\pi}}\,\frac{1}{a^2}
\end{eqnarray}
\subsubsection*{The Relaxation Time Approximation}

Here, the shear viscosity is calculated by combining the results in
Eqs.~(\ref{relax4}) and (\ref{nretasrelax}). For a constant
differential cross section, the collision frequency can be expressed
as \cite{McLennan:89} (see Appendix for details of the derivation)
\begin{eqnarray}
w_a(v_a) &=& n\,\sigma_T\,\sqrt{\frac{2k_BT}{\pi\,m}}\,
 \left[ e^{-\zeta^2}+ \left(2\zeta + \zeta^{-1} \right)\,\int_0^{\zeta}\,dt\,e^{-t^2} \right],\nonumber \\
 \label{collfreq}
\end{eqnarray}
where the dimensionless variable
\begin{equation}
 \zeta = \sqrt{\frac{m}{2k_BT}}\,v\,.
\end{equation}
The shear viscosity, from Eq. (\ref{nretasrelax}), is
\begin{eqnarray}
 \eta_s &=& \frac{1}{30\pi^2}\,\frac{m^5}{T}\,\int\,dv_a\,\frac{v_a^6}{n\,\sigma_T}\,\sqrt{\frac{\pi\,m}{2k_BT}}\,\nonumber \\
&& \times \,
~ \frac{f_a^{eq}}{\left[ e^{-\zeta^2}+ \left(2\zeta + \zeta^{-1} \right)\,\int_0^{\zeta}\,dt\,e^{-t^2} \right]}\,.
\end{eqnarray}
In the non-relativistic limit
\begin{eqnarray}
 f_a^{eq} &=& \exp (\mu/k_BT)~  e^{-\zeta^2} \,\,,
\end{eqnarray}
where $\mu$ is the chemical potential.
Hence
\begin{eqnarray}
 \eta_s &=&\frac{1}{30\pi^2}\,\frac{m^5}{T}\,\frac{z}{n\,\sigma_T}\,\sqrt{\frac{\pi\,m}{2k_BT}}\,~\nonumber \\
&& \times\, \int\,dv_a\,v_a^6\,\frac{e^{-\zeta^2}}{\left[ e^{-\zeta^2}+ \left(2\zeta + \zeta^{-1} \right)\,\int_0^{\zeta}\,dt\,e^{-t^2} \right]}\,.
\end{eqnarray}
A numerical quadrature of the above integral yields the result $0.463282\,\left(\frac{2k_BT}{m} \right)^{7/2} $.
The shear viscosity is then given by
\begin{eqnarray}
 \eta_s &=& \frac{3.706256\,\sqrt{\pi}}{30\pi^2}\,\frac{\exp (\mu/k_BT)}{n\,\sigma_T}\,\frac{m^2c^4\,k_B^2T^2}{\hbar^3 c^3}\,\,,
\end{eqnarray}
where the factor of $(\hbar c)^3$ has been inserted to get the correct unit of viscosity.
For Bolztmann statistics
\begin{eqnarray}
 n &=&  \exp(\mu/k_BT)\,\left(\frac{mc^2\,k_BT}{2\pi\hbar^2c^2} \right)^{3/2}\,\,,
\label{ndensity}
\end{eqnarray}
so that
\begin{eqnarray}
 \eta_s &=&  \frac{0.34942}{4 \sqrt{\pi}}\,\sqrt{\frac{m\,k_BT}{\pi}}\,\frac{1}{a^2}
\end{eqnarray}

\subsection{The Hard Sphere Gas (Ultra-Relativistic)}
In this section, we calculate the shear viscosity for massless particles using
a constant differential cross section $\sigma_0 = a^2/4$ for which the total
cross section $\sigma_T = \pi \, a^2$, where $a$ is the radius of the sphere. For point particles such as quarks, the quantity $a$ can be regarded as an effective length scale that serves to define a scattering cross section.

\subsection*{The Chapman-Enskog Approximation}
We start by simplifying the omega integral in Eq. (\ref{omegamassless}):
\begin{eqnarray}
 w_{ij}^k &\equiv& \frac{\pi}{2^4\beta}\,\int_{-1}^1\,d\cos\Theta\,\int_0^{\infty}\,
dP\,\sigma(\Theta,P)\,(1-\cos^k\Theta) \, \nonumber \\
&& \times\, (\beta c P)^i\,K_i(\beta c P) \,\nonumber.
\end{eqnarray}
For $s = 2 $ and a constant cross section $\sigma(\Theta, P) = \sigma_0$, we can rewrite the above
equation as
\begin{eqnarray}
 w_{\mu\nu}^2 = \frac{\pi\,\sigma_0}{12\,\beta}\,\int_0^{\infty}\,dp\,(\beta c p)^{\mu}
\,K_{\nu}(\beta c p)\,.
\end{eqnarray}
The integration can be performed through a change of variable :
\begin{eqnarray}
 \beta c p = 2x\quad \Rightarrow \quad dp = \frac{2}{\beta c}\, dx\,\,.
\end{eqnarray}
Then the omega integral can be written as
\begin{eqnarray}
 w_{\mu\nu}^2 = \frac{\pi\,\sigma_0}{\beta^2c}\,\frac{2^{\mu-1}}{3}\,
\int_0^{\infty}\,dx\,x^{\mu}\,K_{\nu}(2x)\,.
\end{eqnarray}
We can now use the identity
\begin{eqnarray}
 \int_0^{\infty}\,dx\,x^{\mu}\,K_{\nu}(2x) = \frac{1}{4}\,\Gamma\,\left[
(\mu+\nu+1)/2 \right]\,\Gamma\left[ (\mu-\nu+1)/2 \right]\,\,\nonumber\\
\end{eqnarray}
to write the omega integral for massless particles with a constant cross section as
\begin{eqnarray}
 w_{\mu\nu}^2 =\frac{\pi\,\sigma_0}{\beta^2c}\,\frac{2^{\mu-1}}{12}\,\Gamma\,\left[
(\mu+\nu+1)/2 \right]\,\Gamma\left[ (\mu-\nu+1)/2 \right]\,\,.\nonumber \\
\end{eqnarray}
The shear viscosity is
\begin{eqnarray}
 \eta_s = -4nc^{-2}\,(k_BT)^3\,c_0\,\,,
\label{shearmassless2}
\end{eqnarray}
where $c_0$ can be calculated from Eq. (\ref{sumofcnlinear}). For the first order calculation,
the coefficient $c_0$ is
\begin{eqnarray}
 \left[ c_0 \right]_1 = -\frac{40}{cn^2}\,(k_BT)^2\,\frac{1}{c_{00}}\,\,.
\end{eqnarray}
Substituting for $c_0$ in Eq. (\ref{shearmassless2}),
\begin{eqnarray}
 \eta_s = \frac{160}{c^4}\,\frac{1}{\beta^5}\,\frac{1}{c_{00}}
\end{eqnarray}
where $c_{00}$ is taken from Eq. (\ref{c00massless}). The omega integrals needed are
\begin{eqnarray}
 w_{63}^2 = 64\,\frac{\pi\,\sigma_0}{\beta^2c}\,,\,\,\,
w_{72}^2 = 256\,\frac{\pi\,\sigma_0}{\beta^2c} \,\, {\rm and} \,\,
w_{81}^2 = 1536\,\frac{\pi\,\sigma_0}{\beta^2c}\,\,. \nonumber \\
\end{eqnarray}
Utilizing these results, the first order approximation for the
shear viscosity is
\begin{eqnarray}
 [\eta_s]_1 = 1.2\,\frac{k_BT}{\pi\,a^2}\,\frac{1}{c}\,\,,
\end{eqnarray}
where we have used $\sigma_0 = a^2/4$, where $a$ is the radius of the
hard sphere.
\subsubsection*{Successive Approximations}

\begin{figure}
\includegraphics[width=9cm]{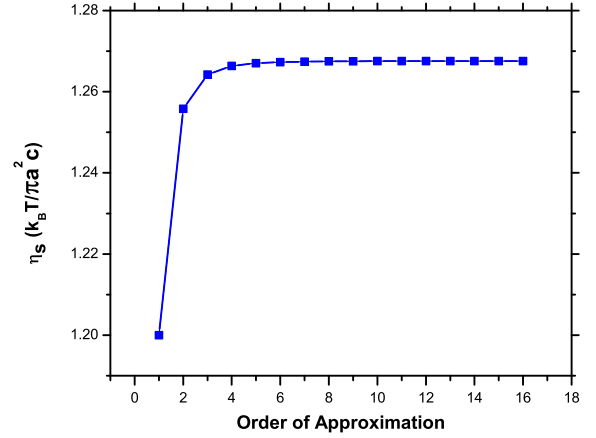}
\caption{The shear viscosity calculated up to the 16th order approximation. }
\label{masslessconstant}
\end{figure}

A point worth noting is that succcessive approximations to the shear
viscosity can be obtained in the Chapman-Enskog approximation, a
feature that is lacking in the relaxation time approximation. As an
example, we calculated up to 16 orders in the case of the
ultra-relativistic hard sphere gas.
For higher order calculations, the nested sums in Eqs. (\ref{cmns}) and
(\ref{cmnis}) in the calculation of the coefficients $c_{mn}$ and
$c_{mn,i}$ call for a large number of evaluations (many repeated) of the omega integrals.
Fortunately, the omega integrals required
in this calculation can be performed analytically (consuming little computer time):
\begin{eqnarray}
{\tilde \omega}_{ij}^k = \frac{\pi\,a^2}{\beta\,c}\,\frac{2^i}{2^6}\,\left(\frac{i+j-1}{2} \right)!
\left(\frac{i-j-1}{2} \right)! \left( \frac{k}{k+1} \right) \,\,\,\,\,\,\,\,.
\end{eqnarray}
Our results for shear viscosity  are shown in the second column of Table \ref{masslessviscosityconstant}. and in Fig. \ref{masslessconstant}. We note that results up to the third order approximation exist in the literature in Ref. \cite{Kox:76}.  Our test of the convergence of higher order approximations here indicate that for all practical purposes the
third order results are adequate. In addition, $z=mc^2/k_BT$ corrections are also available for the third order  results, which can be gainfully employed to check results of computer simulations in which the mass cannot strictly be set to zero.

\begin{table}[h]%The best place to locate the table environment is directly after its first reference in text
\caption{\label{masslessviscosityconstant}Shear viscosity for the ultra-relativistic hard shpere gas up to the
16th order.}\begin{ruledtabular}
\begin{tabular}{cc}
Order of approximation & $ \eta_s  / \,[ k_BT/(\pi a^2~c)]$  \\
\hline\hline
 1&  1.2 \\
 2&  1.25581395 \\
 3&  1.2642487  \\
 4&  1.2663424 \\
 5 & 1.26703133 \\
 6 & 1.26730375 \\
 7 & 1.26742645 \\
 8 & 1.26748735 \\
 9 & 1.26751995 \\
 10 & 1.26753849 \\
 11&  1.26754958  \\
 12 & 1.26755648 \\
 13 & 1.26756094 \\
 14 & 1.26756391  \\
 15 & 1.26756593 \\
 16 & 1.26756759 \\
\end{tabular}
\end{ruledtabular}
\end{table}

In addition to showing the convergence of results, the final result in
this case serves as a test-bed result that Green-Kubo calculations
can shoot for. Such calculations are underway
and will be reported elsewhere.

\subsection*{The Relaxation Time Approximation}
Here, we start with the collision frequency in Eq. (\ref{freqrelax2}).
For a constant cross section
\begin{eqnarray}
 w_a(E_a) &=& \frac{1}{16\,\pi^2}\,\frac{\sigma_T}{E_a}\,
\int_{0}^{\infty}\,dE_b\,\int_{-1}^{1}\,dx\,E_b\,s\,e^{-E_b/k_BT}\,,\nonumber \\
\end{eqnarray}
where $x=\cos \theta$.
To solve the above integral, we note that
\begin{eqnarray}
 s - (p_a+p_b)^2 = 2\,|p_a|\,|p_b|\,\left(x - \frac{s}{2|p_a|\,|p_b|}
\right)\,.
\end{eqnarray}
Inserting the identity involving the delta function
\begin{eqnarray}
 1 = \int\,ds\,\frac{1}{2\,|p_a|\,|p_b|}\,\delta\left(x - \frac{s}{2|p_a|\,|p_b|} \right)
\end{eqnarray}
in the above integral, the collision frequency reads as
\begin{eqnarray}
w_a(E_a) &=& \frac{\sigma_T\,(k_BT)^3}{2\,\pi^2}\,\,.
\end{eqnarray}
Supplying the collision frequency into
Eq. (\ref{etasrelax}), the shear viscosity becomes
\begin{eqnarray}
 \eta_s = \frac{8}{5}\,\frac{k_BT}{\sigma_T}\,\frac{1}{c}\,.
\end{eqnarray}

If we use the Bose-Einstein distribution function in the calculation, then the result for
the collision frequency is given by
\begin{eqnarray}
 w_a(E_a) = \frac{\sigma_T\,(k_BT)^3}{2\,\pi^2}\,\zeta(3)
\end{eqnarray}
and the shear viscosity is given by
\begin{eqnarray}
 \eta_s = \frac{8}{5}\,\frac{k_BT}{\sigma_T}\,\frac{1}{c}\,\frac{\zeta(5)}{\zeta(3)}\,,
\end{eqnarray}
where $\sigma_T = \pi\,a^2$.
\subsection{The Maxwell Gas}
Particles in the Maxwell gas are characterized by the differential cross section~\cite{Polak:73}
\begin{eqnarray}
 \sigma(g, \theta) = \frac{m\,\Gamma(\theta)}{2\,g}\,\,,
\end{eqnarray}
where $m$ is the mass, $g$ is the relative momentum and
$\Gamma(\theta)$ is an arbitrary function of angle.
The unit of $\Gamma(\theta)$ is fm$^3/$s. In this calculation,
we set $\Gamma(\theta) = \Gamma $, where $\Gamma$ is   a constant.
Inclusion of $\theta-$ dependence is straighfoward.
\subsubsection*{The Chapman - Enskog Approximation}
We begin by writing the cross section in terms of  $\phi = g/\sqrt{mT}$ as
\begin{eqnarray}
 \sigma(g,\theta) &=& \frac{m}{2g}\,\Gamma(\theta) = \frac{m\,\Gamma(\theta)}{2\,\phi\,\sqrt{mT}}
\end{eqnarray}
The non relativistic omega integral for Maxwell particles can be calculated by using Eq.
(\ref{nrelomega}) :
\begin{eqnarray}
\Omega_2^{(2)}  = \frac{5}{4}\,\pi\,\Gamma
\end{eqnarray}
The quantity $c_{00}$ can be calculated from Eq. (\ref{c00massive}) with the result
 $ c_{00} = 20\,\pi\,\Gamma $.
From Eq. (\ref{shearfirst}), the shear viscosity is
\begin{eqnarray}
 \eta_s &=& \frac{k_BT}{2\,\pi\,\Gamma}\,,
\end{eqnarray}
where we have used $\gamma_0 = -10\hat{h} $ with $\hat{h} \rightarrow 1 $ in the non relativistic limit.
\subsubsection*{The Relaxation Time Approximation}
In the relaxation time approximation, the shear viscosity can be calculated using Eqs.~(\ref{relax4}) and (\ref{nretasrelax}).
We start by calculating the collision frequency
\begin{eqnarray}
 w_a(v_a) &=& \int\,d^3v_b\,d\Omega\,\frac{m\,\Gamma(\theta)}{2m\,|\vec{v}_a-\vec{v}_b|}\,f_b\,\,|\vec{v}_a-\vec{v}_b| \nonumber \\
\end{eqnarray}
For angle independent $\Gamma$, the collision frequency, $ w_a(v_a)  = 2\pi\,\Gamma\,n $.
From Eq. (\ref{nretasrelax}),
\begin{eqnarray}
 \eta_s &=& \frac{1}{30\pi^2}\,\frac{m^5}{T}\,\frac{1}{2\,\pi\,\Gamma\,n}\,z\,\int_0^{\infty}dv_a\,v_a^6\,e^{-mv_a^2/(2k_BT)}\,\,\,.
\end{eqnarray}
The quantity $n = z\left(\frac{mT}{2\pi\hbar^2c^2} \right)^{3/2}$ is defined as in Eq.
(\ref{ndensity}). Setting
\begin{equation}
 x = \sqrt{\frac{m}{2k_BT}} \, v_a
\end{equation}
and performing the integral,
the shear viscosity is
\begin{eqnarray}
 \eta_s &=&  \frac{k_BT}{2\pi\Gamma}\,.
\end{eqnarray}

\subsection{Massless Pions}
In the relativistic regime, we consider massless pions whose elastic differential cross section
is given by \cite{PPVW:93},
\begin{eqnarray}
 \sigma(s,\theta) = \frac{s}{64\,\pi^2\,f_{\pi}^4}\,\,\left(3 + \cos^2\theta \right)
\end{eqnarray}
where $s = \sqrt{E_{cm}}$ is a Mandelstahm  variable
and $f_{\pi} = 93 $ MeV is the pion decay consntant.
\subsubsection*{The Chapman-Enskog Approximation}
In the Chapman-Enskog approximation, the formalism to calculate the shear viscosity
is described in Ref.~\cite{deGroot:75} and was summarized in section IIB.
The shear viscosity is obtained from Eqs.(\ref{shearmassless}, \ref{cmnis}, \ref{omegamassless} and \ref{c00massless}).
The omega integrals that are required for chiral pions are
\begin{eqnarray}
 \tilde{w}_{63}^2 &=& \frac{128}{\pi\,c^3}\,
\left(\frac{k_BT}{f_{\pi}} \right)^4\,,\,\quad
\tilde{w}_{72}^2 = \frac{768}{\pi\,c^3}\,\left(\frac{k_BT}{f_{\pi}} \right)^4\,\,\,\nonumber \\
&& \,\textnormal{and}\,\,\,\tilde{w}_{81}^2 = \frac{6144}{\pi\,c^3}\,\left(\frac{k_BT}{f_{\pi}} \right)^4
\label{omegaformasslesspion}
\end{eqnarray}
Using the above omega integrals  to calculate $c_{00}$,
we finally obtain the shear viscosity as
\begin{eqnarray}
 \eta_s &=& \frac{15\,\pi}{184}\,\frac{f_{\pi}^4}{k_{B}T}\,\frac{1}{c} .
\end{eqnarray}
\subsubsection*{The Relaxation Time Approximation}
We first calculate the collision frequency  defined in Eq. (\ref{freqrelax2}).
In order to perform the integral we set $x = \cos\,\theta$ and  note that
\begin{eqnarray}
 s - (k_a+k_b)^2 &=& 2|k_a|\,|k_b|\,\left(x - \frac{s}{2|k_a|\,|k_b|} \right)
\end{eqnarray}
Introducing the identity involving the delta function
\begin{eqnarray}
  1 &=& \int\,ds\,\, \frac{1}{2|p_a|\,|p_b|}\,\,\delta\left(x - \frac{s}{2|p_a|\,|p_b|} \right)\,\,,
\end{eqnarray}
and inserting the above two relations in  Eq. (\ref{freqrelax2}),
we arrive at
\begin{eqnarray}
  w_a(E_a) &=& \frac{\pi}{108}\,E_a\,\left(\frac{k_BT}{f_{\pi}} \right)^4 \,.
\end{eqnarray}
Inserting the above result into  Eq. (\ref{etasrelax}), the shear viscosity
for chiral pions reads as
\begin{eqnarray}
 \eta &=&\frac{12\pi}{25}\,\frac{f_{\pi}^4}{T}\frac {1}{\hbar^2c^3}\,.
\end{eqnarray}

\subsection{Interacting Massive Pions}

We choose the following parameterization for the experimental $\pi - \pi$ phase
shifts adopted by Bertsch et al.,~\cite{Bertsch:88}:
\begin{equation}
 \delta_0^0 = \frac{\pi}{2} + \arctan\left( \frac{\epsilon-m_{\sigma}}{\Gamma_{\sigma} /2 }\right)
\end{equation}
\begin{equation}
 \delta_1^1 = \frac{\pi}{2} + \arctan\left( \frac{\epsilon-m_{\rho}}{\Gamma_{\rho} /2 }\right)
\end{equation}
\begin{equation}
 \delta_0^2 = -\frac{0.12 q}{m_{\pi}},
\end{equation}
where in the symbol $\delta_l^I$,
$I$ is the total isospin of the two pions and $l$ is the angular momentum. The quantity
\begin{equation}
 \epsilon = 2\left( q^2 + m_{\pi}^2\right) ^{1/2} \qquad {\rm with} \qquad m_{\pi} = 140 \textnormal{ MeV}
\end{equation}
The phase shift $\delta_0^0$ corresponds to the $s$-wave $\sigma$ resonance,
with the width $\Gamma_{\sigma} = 2.06~ q$ and $m_{\sigma} = 5.8~ m_{\pi}$.
The phase shift $\delta_1^1$ is from the $p$-wave $\rho$ resonance, with the width
\begin{equation}
 \Gamma_{\rho}(q) = 0.095 q \left( \frac{q/m_{\pi}}{1+\left( q/m_{\rho}\right)^2 }\right)^2
\end{equation}
and $m_{\rho} = 5.53 ~m_{\pi}$. The phase shift $\delta_0^2$ accounts for $s$-wave repulsive interactions.
The isospin averaged cross section for elastic scattering is obtained from
\begin{equation}
 \frac{d\sigma(q,\Theta)}{d\Omega} = \frac{4}{q^2} \sum_{l,I}^{'}
\frac{(2I+1)(2l+1)}{\sum_{I}(2I+1)} \,P_l(\cos \Theta)\, \sin^2\delta_l^I(q),
\end{equation}
where the prime denotes that the isospin sum is restricted to values for which $l + I$ is even, and $l = 0,1,2 $, whence
\begin{equation}
\frac{d\sigma(s,\Theta)}{d\Omega} = \frac{4}{q_{cm}^2}\left( \frac{1}{9}\sin^2\delta_0^0+
\frac{5}{9}\sin^2\delta_0^2+\frac{1}{3}~9\sin^2\delta_1^1\cos^2\Theta\right)\,.
\label{diffcross}
\end{equation}

\begin{figure}[tbp]
\includegraphics[width=8cm]{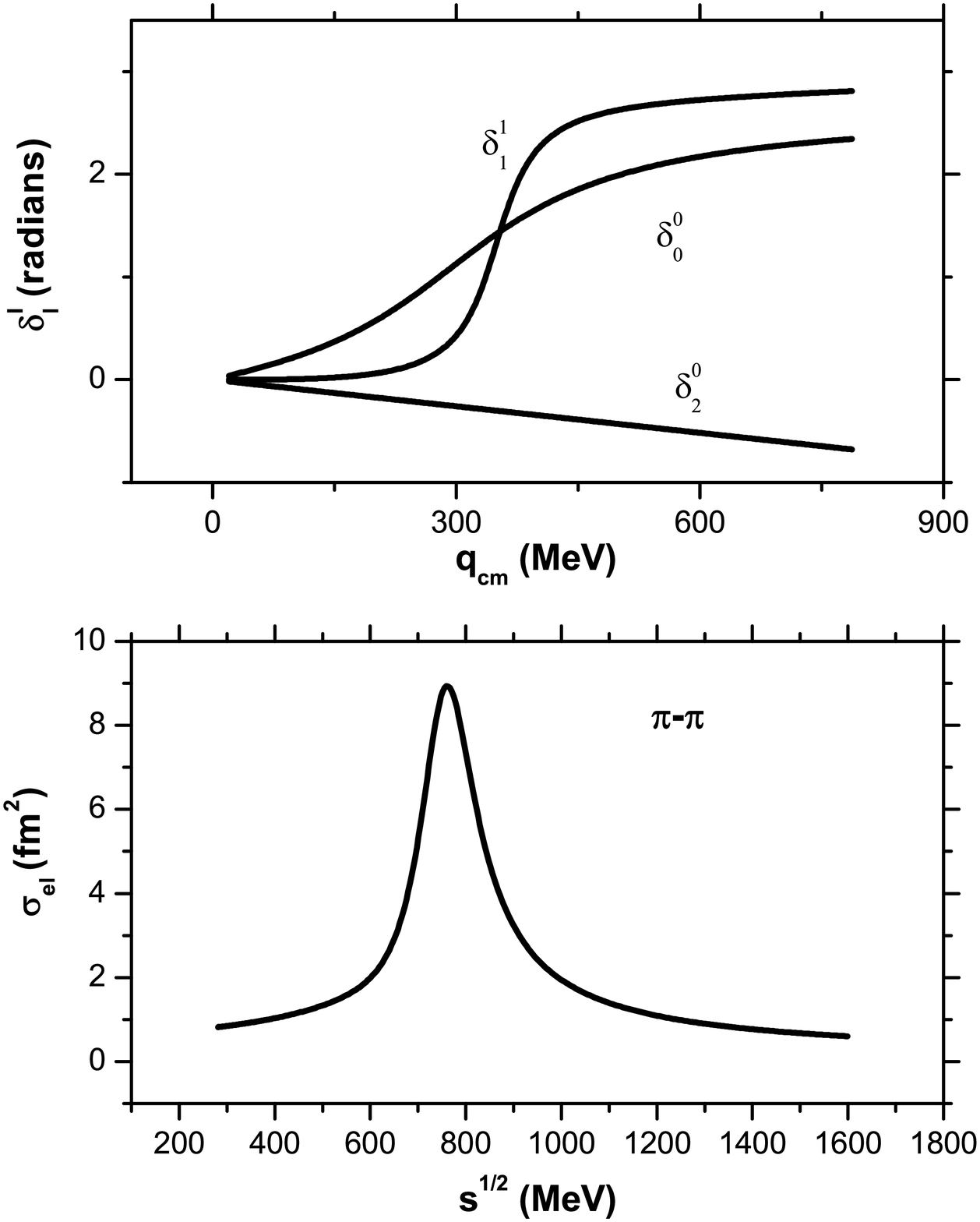}
\caption{Top panel: Contributing phase shifts for pion-pion
scattering.
The phase shift $\delta_0^0$ corresponds to the $s$-wave $\sigma$ resonance,
$\delta_1^1$ is from the $p$-wave $\rho$ resonance, and
$\delta_0^2$ accounts for $s$-wave repulsive interactions.
Bottom panel: Total cross section versus the center of mass energy. This figure is adapted from Venugopalan and Prakash, Nucl. Phys. A {\bf 546} (1992) 718.}
\label{crosssection}
\end{figure}
In Fig.~\ref{crosssection}, the phase shifts (top panel) and the total cross section (bottom panel) are shown. The $s$-wave $\sigma$ resonance and the $p$-wave $\rho$ resonance
are clearly evident from this figure as the corresponding phase shifts $\delta_0^0$ and
$\delta_1^1$ both exceed $\pi$ radians, a signature of resonance  formation. Note that the total cross section is dominated by the $\rho$ resonance with a peak around 770 MeV.

\begin{figure}[tbp]
\includegraphics[width=9cm]{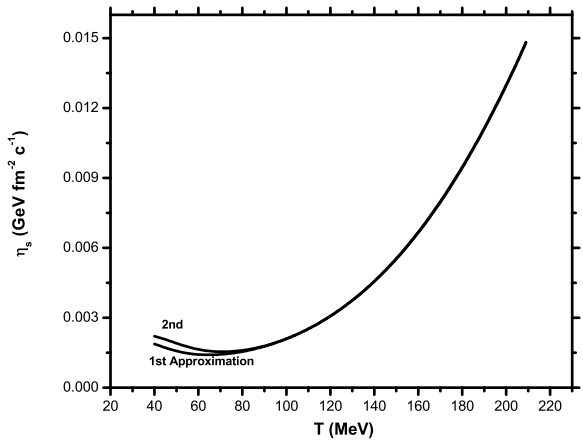}
\caption{Shear viscosity versus temperature for a system of interacting pions with experimental cross sections.
Results up to the second order approximation are shown. This figure is adapted from Prakash et al., Phys. Rep. {\bf 227} (1993) 331.}
\label{shearforpion}
\end{figure}

The results for the shear viscosity of interacting pions up to the second order
approximation are shown in Fig. \ref{shearforpion} using the experimental
differential cross sections.
For calculating results beyond the first order approximation, methods described
in Sec. \ref{Formalism} are employed. The role of the energy dependence of
the scattering cross section is evident from this figure. Beyond the $\rho$-
meson resonance energy of 770 MeV, the experimental cross sections decrease
with the center of mass energy which makes the shear viscosity increase  with
temperature. The results also show the rapid convergence of the Chapman-Enskog
approach for the shear viscosity. The first order results appear quite adequate
for all practical purposes in the temperature range of 100-200 MeV of relevance to heavy-ion collisions.

\begin{figure}[tbp]
\includegraphics[width=9cm]{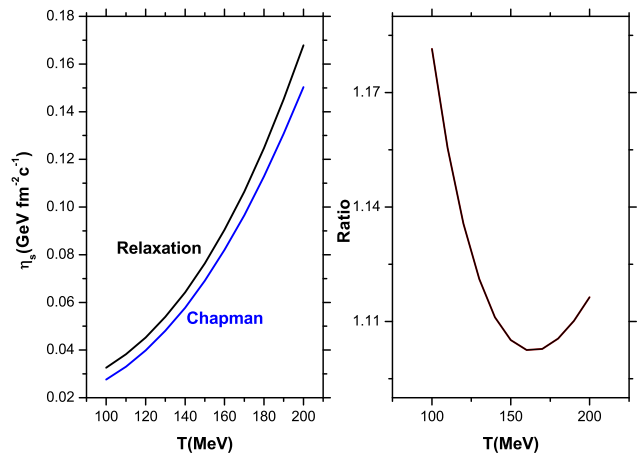}
\caption{Left panel: Shear viscosities of pions gas from the
relaxation time approximation
and the first order Chapman-Enskog approximation.
Right panel: The ratio of the results in the left panel.}
\label{piongascomparison}
\end{figure}

In Fig. \ref{piongascomparison}, the first order results of shear viscosity
from the Chapman-Enskog approach are compared with those from the relaxation
time approach (left panel).
The right panel shows the ratio which is calculated as the result from the
Chapman-Enskog viscosity divided the result by the Relaxation time viscosity.

\begin{table*}
\caption{Summary of results for shear viscosity.
Results for the Chapman-Enskog approach are for the first order approximation.
\label{tab:table2}}
\begin{tabular}{lcccc}
\hline
Case & Cross-section & Chapman-Enskog & Relaxation & Ratio \\
\hline \hline
& & & & \\
&  \multicolumn{2}{c}{Shear viscosities of nonrelativistic systems}  & & \\
& & & & \\
Hard-sphere (Nonrelativistic) & $\sigma = \frac{a^2}{4}$ & $0.078\,
\sqrt{\frac{m\,k_BT}{\pi}}\,\frac{1}{a^2}$ &
$0.049\,\sqrt{\frac{m\,k_BT}{\pi}}\,\frac{1}{a^2}$ &1.59\\
& & & & \\
Maxwell gas & $\sigma_0 = \frac{m\,\Gamma(\theta)}{2\,g}$ & $\frac{k_BT}{2\,\pi\,\Gamma}$
& $\frac{k_BT}{2\pi\,\Gamma} $ &1.00 \\
& & & & \\
&  \multicolumn{2}{c}{Shear viscosities of ultrarelativistic systems}  & & \\
& & & & \\
Hard-sphere (Ultrarelativistic) & $\sigma_0 = \frac{a^2}{4}$ & $ 1.2\,\frac{k_BT}{\pi\,a^2}\,\frac{1}{c}$ &  $ \frac{8}{5}\,\frac{k_BT}{\pi\,a^2\,c}\,$ & 0.75  \\
& & & & \\
Chiral pions & $\sigma = \frac
{s}{(64\pi^2 f_\pi^4)} \left(3 + \cos^2 \theta \right)$ & $\frac{15\pi}{184}\,\frac{f_{\pi}^4}{T} \frac {1}{\hbar^2c^3}$ & $\frac{12\pi}{25}\,\frac{f_{\pi}^4}{T}\frac {1}{\hbar^2c^3} $ &0.169 \\
& & & & \\
\hline
\end{tabular}
\end{table*}
\section{Discussion of Analytical and Numerical Results}
\label{Discussion}

In this section, we collect results of calculations performed using
the two different approaches, the Chapman-Enskog approximation and
the relaxation time approximation. The non relativistic limit ($z =
mc^2/k_BT \gg 1$) is examined in the cases of the hard sphere particles
(non-relativistic case) and the Maxwell particles.  The
ultra-relativistic limit is explored in the cases of the hard sphere
gas and massless pions. In the case of massive interacting pions with
experimental cross sections, calculations are performed using the
general relativistic scheme outlined in Sec. \ref{Formalism}.

Table \ref{tab:table2} shows the systems considered along with with their
corresponding cross sections, and results of $\eta_s$ from the
first order Chapman-Enskog  and the relaxation time approximations.
The results in the table  and those in
the following figures must be viewed bearing in mind one
difference that exists in the calculational procedures. The
Chapman-Enskog approximation features the transport cross section with
an angular weight of $(1 - \cos^2 \Theta)$ in first order
calculations. The relaxation time approach lacks this angular
weighting. The angular integral can be performed analytically for the
cases chosen and leads to a factor of 4/3 for angle independent cross sections.
Even so, it is intriguing
that for the case of Maxwell particles, the two methods give exactly
the same result. This is perhaps because of the fact that the relative
velocity appearing in the denominator of the cross section is exactly
cancelled by a similar factor occuring in the numerator in both
methods. In the remaining cases, it is clear from Table \ref{tab:table2} that the
energy dependence of the cross sections plays a crucial role in
determining the extent to which results differ between the two
approaches.
%

%\clearpage

\section{Summary and Conclusions}
\label{Summary}

A quantitative comparison between results from the
Chapman-Enskog and relaxation time methods to calculate viscosities
was undertaken for the following test cases:

\begin{enumerate}

\item The non-relativistic and relativistic hard sphere gas in which particles interact with a constant cross section;

\item  The Maxwell gas in which the cross section is inversely proportional to the relative velocity of the scattering particles;

\item Chiral pions for which the cross section is proportional to the squared center of mass energy; and

\item Massive pions for which the differential elastic cross section features resonances
is taken from experiments.

\end{enumerate}

The analytical and numerical results of our comparative study reveal that
the extent of agreement (or disagreement) depends sensitively on the energy
dependence of the differential cross sections employed.  Our calculations of
the shear viscosity of ultra-relativistic hard spheres can be used to check
Green-Kubo calculations of shear viscosity, a test that is being undertaken currently.

\section*{Acknowledgements}

We thank J. I. Kapusta and S. Gavin for helpful discussions and communications.
We are grateful to Purnendu  Chakraborty  for providing us results of the viscosity of massive pions in the relaxation time method.
Research support from U. S. DOE grants DE-AC02-05CH11231 and Central China Normal University through
colleges of basic research \& operation of MOE (for A. W.),  from the U. S. DOE grants DE-FG02-93ER-40756 (for A. W. and M. P.) and DE-FG02-87ER40328 (for P. C.) are gratefully acknowledged.

\section*{Appendix}
\subsection*{Nonrelativistic Collision Frequency (Eq. (\ref{collfreq}))}
\label{frequency}
We start from the expresion
\begin{eqnarray}
 w_a(v_a) &=& \int_0^{\infty} \, d^3v_b\,\sigma_T\,f_b\,|\vec{v}_a - \vec{v}_b|\,\,,
\end{eqnarray}
where
\begin{eqnarray}
 f_b = n\,\left(\frac{m}{2\pi\,kT} \right)^{3/2}\,e^{-mv_b^2/(2kT)}\,.
\end{eqnarray}
For a constant differential cross section, the collision frequency is
\begin{eqnarray}
 w_a(v_a) &=& \sigma_T\,n\,\left(\frac{m}{2\pi\,kT} \right)^{3/2}\,2\pi\,\int\,
 dv_b\,v_b^2\,e^{-mv_b^2/(2kT)}\,\nonumber \\
&& \times\, \int_{-1}^{1}\,|\vec{v}_a-\vec{v}_b|
\end{eqnarray}
Setting $\sqrt{\frac{m}{2kT}}\,v_b = \zeta_b\,\,,$ we have
\begin{eqnarray}
w_a(v_a) &=& 2\,\sigma_T\,n\,\left(\frac{2kT}{m} \right)^{1/2}\,
\int_0^{\infty}\,d\zeta_b\,\,\zeta_b^2\,\,e^{-\zeta_b^2}\,\nonumber \\
&& \times\,\int_{-1}^{1}\,dx\,|\vec{\zeta}_a-\vec{\zeta}_b|
\end{eqnarray}
The angular integration yields
\begin{eqnarray}
 \int_{-1}^{1}\,dx\,|\vec{\zeta}_a-\vec{\zeta}_b| =  \frac{1}{3\,\zeta_a\,\zeta_b}\,
 \left((\zeta_a+\zeta_b)^3 - (\zeta_a-\zeta_b)^3 \right)\,\,\,\,\, \nonumber \\
\label{eq65}
\end{eqnarray}
The above expression can be further simplified in the cases
\begin{eqnarray}
\zeta_a > \zeta_b\quad \textnormal{for which}\quad \int_{-1}^{1}\,dx\,|\vec{\zeta}_a-\vec{\zeta}_b| =  2\,\left(\zeta_a + \frac{\zeta_b^2}{3\,\zeta_a} \right) \nonumber \\
\zeta_a < \zeta_b \quad \textnormal{for which}\quad \int_{-1}^{1}\,dx\,|\vec{\zeta}_a-\vec{\zeta}_b| = 2\,\left(\zeta_b + \frac{\zeta_a^2}{3\,\zeta_b} \right)\,.\nonumber \\
\end{eqnarray}
These two expressions can be inserted into Eq. (\ref{eq65}) so that the integration over $\zeta_b$ reads as
\begin{eqnarray}
 2\,\left(\int_0^{\zeta_a}d\zeta_b\,\,\zeta_b^2\,e^{-\zeta_b^2}
\left(\zeta_a + \frac{\zeta_b^2}{\zeta_a} \right) \right.  \nonumber \\
 \left. +
\int_{\zeta_a}^{\infty}d\zeta_b\,\,\zeta_b^2\,e^{-\zeta_b^2} \left(\zeta_b +
\frac{\zeta_a^2}{\zeta_b} \right) \right)
\end{eqnarray}
Upon integration by parts,
the first term of the above integration results in two terms:
\begin{eqnarray}
 \zeta_a\,\int_0^{\zeta_a} \,d\zeta_b\,\,\zeta_b^2\,\,e^{-\zeta^2} &=&
 - \frac{1}{2}\,\zeta_a\,e^{-\zeta_a^2}  \nonumber \\
 && +\frac{1}{2} \int_0^{\zeta_a} \,d\zeta_b\,e^{-\zeta_b^2}
\end{eqnarray}
The second term, again integration by parts gives
\begin{eqnarray}
 \frac{1}{\zeta_a}\,\int_0^{\zeta_a} \,d\zeta_b\,\,\zeta_b^3\,\,(\zeta_b\,e^{-\zeta_b^2})
 &=& - \frac{\zeta_a^3}{2}\,e^{-\zeta_a^2} - \frac{3\zeta_a}{4}\,e^{-\zeta_a^2}
 \nonumber \\
 && + \frac{3}{4}\,\int_0^{\zeta_a} d\zeta_b\,\,e^{-\zeta_b^2}
\end{eqnarray}
The sum of these two terms is
\begin{eqnarray}
 I_1 &=& - \frac{4}{3}\,\zeta_a^2\,\,e^{-\zeta_a^2} \nonumber \\ \
 && + \left(\zeta_a + \frac{1}{2\zeta_a} \right)\,\int_0^{\zeta_a}\,d\zeta_b\,\,e^{-\zeta_b^2}-\frac{1}{2}\,\,e^{-\zeta_a^2}
\end{eqnarray}
The last two terms in Eq. (\ref{eq65}) can be integrated to yield
\begin{eqnarray}
 I_2 &=& 2 \left( \int_{\zeta_a}^{\infty}\,d\zeta_b\,\zeta_b^3\,e^{-\zeta_b^2} + \frac{\zeta_a^2}{3}\,\int_{\zeta_a}^{\infty}\,d\zeta_b\,\,\zeta_b\,e^{-\zeta_b^2}\right) \nonumber \\
&=& ( 1 + \zeta_a^2 ) \,e^{-\zeta_a^2} + \frac{\zeta_a^2}{3}\,e^{-\zeta_a^2} = ( 1 + \frac{4}{3}\,\zeta_a^2) \,e^{-\zeta_a^2}
\end{eqnarray}
The sum of $I_1$ and $I_2$ is
\begin{eqnarray}
 I_T = I_1 + I_2 = e^{-\zeta_a^2}+ \left(2\zeta_a + \zeta_a^{-1} \right)\,\int_0^{\zeta_a}\,dt\,e^{-t^2}
\end{eqnarray}
Therefore, the collision frequency for the hard sphere gas is given by
\begin{eqnarray}
w_a(v_a) &=& n\,\sigma_T\,\sqrt{\frac{2kT}{\pi\,m}}\,\left[ e^{-\zeta_a^2}+ \left(2\zeta_a + \zeta_a^{-1} \right)\,\int_0^{\zeta_a}\,dt\,e^{-t^2} \right]\,\,. \nonumber \\
\end{eqnarray}
%

\bibliographystyle{h-physrev3}
\bibliography{comparisonreferences}

\end{document}